# Bicrystalline Grain Boundary and Hybrid SNS Junctions Based on Ba-122 Thin Films

Stefan Schmidt, Sebastian Döring, Frank Schmidl, Volker Tympel, Silvia Haindl, Kazumasa Iida, Fritz Kurth, Bernhard Holzapfel, Paul Seidel

*Abstract*—To investigate the properties of iron-based superconductors, we examined different kinds of Josephson junctions using Co-doped Ba-122 thin films. Based on bicrystalline SrTiO$_3$ substrates we prepared grain boundary (GB) junctions, which showed clear Josephson effects. We present electrical measurements at varying temperatures and magnetic fields. Furthermore, we prepared hybrid junctions using thin film technique. Confined by a Ba-122 base electrode (S) and a PbIn counter electrode (S') two different geometries of hybrid junctions are formed: a planar SNS' junction with a gold barrier (N) in between and an edge-type junction with an interface-engineered barrier. All three kinds of junctions showed asymmetric resistively shunted junction-like (RSJ) behavior with $I_C R_N$ products of 20.2 µV (GB junctions), 18.4 µV (planar SNS' junctions) and 12.3 µV (edge-type junctions), respectively. An excess current could be observed at all junctions.

## I. Introduction

JUNCTIONS of superconductors in general give access to basic physical properties of the used materials. Especially, bicrystal grain boundary (GB) junctions are a potent tool for fundamental research. This kind of junctions became well-established for high-$T_C$ superconducting thin film devices, see e.g. the review of Hilgenkamp and Mannhart [1]. For Ba-122 thin films with a thickness of 350 nm on a bicrystalline SrTiO$_3$ (STO)-substrate (with misorientation angles of 3, 6, 9 or 24 degrees) Lee et al. [2] showed that the reduction of the critical current density of small thin film bridges is small compared to the cuprates. Investigations of Katase et al. [3] with Co-doped Ba-122 films on LSAT and MgO bicrystalline substrates with misorientation angles between 3 and 45 degrees demonstrate that the slope of the decrease of the current density on angle is smaller than for the cuprates. Thus, higher critical current densities can be obtained for larger bicrystal angles.

First thin film Josephson junctions with Ba-122 were realized on LSAT bicrystals with grain boundary angle of 30 degrees by Katase et al. [4]. They observed resistively shunted junction-like (RSJ) I-V characteristics up to 17 K, but an $I_C R_N$ of only 56 µV at 4 K with quite linear temperature dependence. With this kind of junctions a dc superconducting quantum interference device (SQUID) was realized by the same group [5] providing voltage-flux modulations of 1.4 µV at 14 K.

In the review of Hiramatsu et al. [6] the conclusion was made that larger junction resistances will be necessary to realize practical SQUIDs with iron-based superconductors. They propose artificial insulating barrier junctions like superconductor-insulator-superconductor (SIS) junctions.

We use well-known bicrystalline STO substrates (angle of 30 degrees) with a Fe buffer layer to improve the quality of the Co-doped Ba-122 thin film.

## II. Junction Preparation

### A. Ba-122 Thin-Film Deposition

Epitaxial BaFe$_{1.84}$Co$_{0.16}$As$_2$ (Co-doped Ba-122) thin films (100 nm) were deposited on Fe buffered (20 nm) [001]-tilt SrTiO$_3$ (STO) bicrystalline substrates with a grain boundary angle ($\theta_{GB}$) of 30° by pulsed laser deposition (PLD). Since epitaxial, smooth Fe buffer layers might be difficult to realize on perovskite substrates (e.g. STO), additional buffer layers of MgAl$_2$O$_4$ (10 nm) were deposited by PLD at 650 °C. Fig. 1(a) shows an example of the reflection high-energy electron diffraction (RHEED) image of epitaxial MgAl$_2$O$_4$ on STO (100) single crystalline substrate. Albeit in a UHV condition

Manuscript received October 9, 2012. This work was partially supported by the DFG within SPP 1458 (project nos. SE 664/15-1 and HA 5934/3-1), the EU within IRON-SEA (project FP7-283141) and the Landesgraduiertenförderung Thüringen.

S. Schmidt, S. Döring, F. Schmidl, V. Tympel, and P. Seidel are with Institut für Festkörperphysik, Low Temperature Physics, Friedrich-Schiller-Universität Jena, Helmholtzweg 5, 07743 Jena, Germany (corresponding author: P. Seidel, +49 3641 947410, paul.seidel@uni-jena.de).

K. Iida, F. Kurth, and B. Holzapfel are with IFW Dresden, Institute for Metallic Materials, Helmholtzstrasse 20, 01069 Dresden, Germany.

S. Haindl is with Leibniz-Institute for Solid State and Materials Research, Institute for Solid State Research, 01069 Dresden, Germany.

Color versions of one or more of the figures in this paper are available online at http://ieeexplore.ieee.org.

Digital Object Identifier 10.1109/TASC.2012.2226861

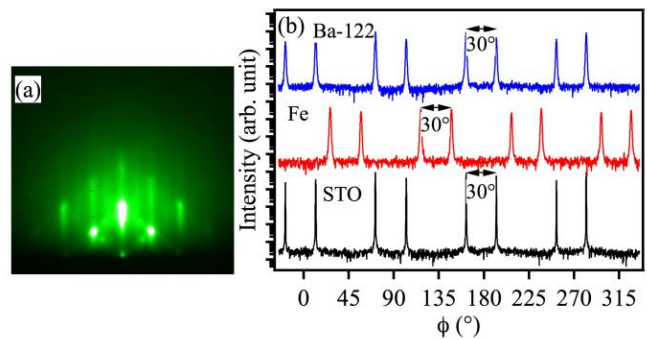

Fig. 1. (a) RHEED image of MgAl$_2$O$_4$ on STO (100) single crystalline substrate. (b) Phi-scans of Co-doped Ba-122, Fe and STO substrate. The respective reflection of Co-doped Ba-122, Fe and STO are (103), (110) and (110).

(base pressure of $10^{-10}$ mbar), $MgAl_2O_4$ can be grown epitaxially on STO substrate. In addition, clear streaks of $MgAl_2O_4$ are visible, indicative of a flat surface. After the primary buffer layer deposition, $MgAl_2O_4$ / STO was cooled to room temperature for the second Fe buffer deposition followed by fabricating Co-doped Ba-122 layers. The details of both Fe buffer and Co-doped Ba-122 depositions are described in [7]. Shown in Fig. 1 (b) are the respective Phi-scans of the (103) Co-doped Ba-122, the (110) Fe, and the (110) STO substrate, measured in a texture goniometer operating with Cu-$K_\alpha$ radiation. A rotation of 45° is visible in Fig. 1(b) between the STO substrate and the Fe layer and also between the Fe layer and the Co-doped Ba-122, reconstituting the alignment to the substrate material. The angular mismatch of the substrate material is passed on to the Fe buffer and the Ba-122 layer, resulting in an artificial grain boundary with $\theta_{GB}$ of 30°.

*B. GB Junctions*

To extract single Josephson junctions we confined the junction area over the grain boundary (GB) by Ion Beam Etching (IBE; 500 eV beam voltage, $10^{-3}$ A·cm$^{-2}$ ion beam density) of narrow bridges (see Fig. 2). The bridge widths range from 4 µm to 20 µm according to our photolithographic mask system. This system has to be adjusted exactly onto the GB, as seen in Fig. 3. Thus, a weak link between the two Ba-122 electrodes is formed in the GB region, which exhibits a Josephson junction. We sputtered a protection layer of gold over the whole sample before the patterning IBE process to avoid unwanted degradation in the photolithography and patterning processes. We can remove the gold layer at the junction area after first measurement in an additional IBE step, because it shunts the junction resistively.

*C. Hybrid Junction Designs*

In contrast to the GB junction type, the hybrid junctions consist of different superconducting materials. The main examination object is the $BaFe_{1.8}Co_{0.2}As_2$ base electrode. As counter electrode we use the conventional superconductor lead. The first hybrid type presented here is the planar SNS' junction, where the weak link between the pnictide base electrode (S) and the counter electrode (S') is formed by a normal conducting barrier layer (N) of gold with a thickness of 5 nm (see Fig. 4). The gold layer can be deposited in-situ after the deposition process of the Co-doped Ba-122 layer, which ensures a clean interface to the pnictide. But being a normal conductor the gold barrier is low ohmic and thus will lead to a junction with a small $I_C R_N$ product. We want to mention, that other normal metals or insulators could be applied ex-situ, too. In this planar sandwich geometry transport processes along the *c*-axis of the Ba-122 thin film are investigated. Reference [6] shows the process in more detail.

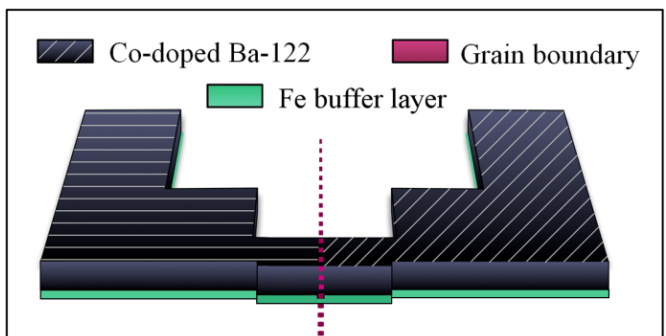

Fig. 2. Schematic layout of a grain boundary (GB) junction. The angular mismatch of a bicristalline substrate is passed on through the Fe buffer layer to the Co-doped Ba-122 thin film. A narrow bridge over the GB area forms a weak link between the two superconducting electrodes on the left and on the right.

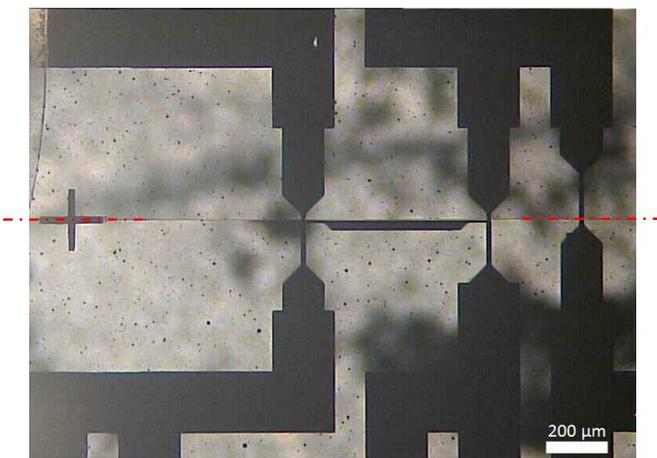

Fig. 3. Transmitted light picture of three bridges over a grain boundary. The GB can be seen as a thin line in the middle, oriented horizontally (marked by the dashed lines). The dark dots and clouds within the bright area are impurities and dirt effects from underneath the substrate material.

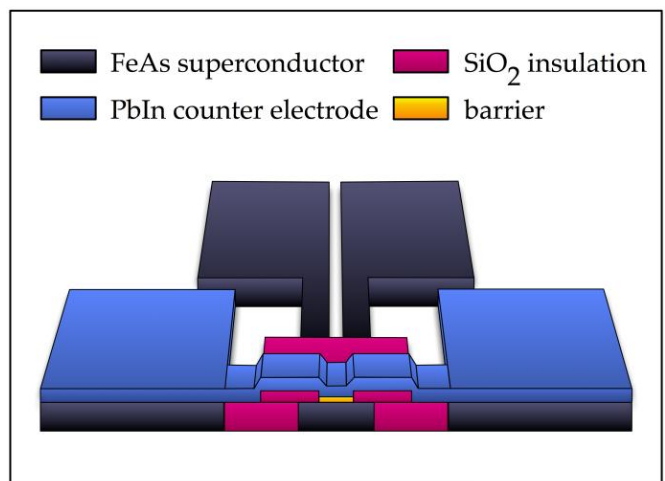

Fig. 4. Schematic layout of a planar SNS' junction. The weak link is realized by a normal conducting barrier (gold) of a given thickness (5 nm) and area (between 3 x 3 µm² and 100 x 100 µm²).

The second hybrid type is the so called edge-type junction (see Fig. 5). The charge carriers are transported through the interface between the two electrodes. This interface is influenced by several photolithography and IBE processes. Contrary to the planar junction type, information about the transport within the *ab*-plane of the pnictide superconductor can be accessed by the edge-type junction. Details on the preparation can be found in [8].

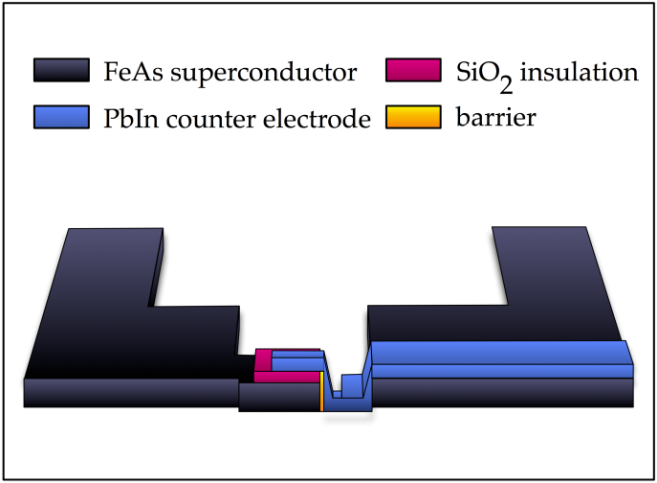

Fig. 5. Schematic layout of an edge-type junction. The interface between both electrodes was influenced by the preparation processes and forms thus a non-superconducting barrier. The junction area can be tuned by the width of the PbIn counter electrode (between 3 µm and 20 µm).

### III. MEASUREMENT RESULTS

#### A. I-V Characteristics and the Excess Current

The current-voltage characteristics of the grain boundary junction show nonlinear behavior at 4.2 K, which can be described with the Resistively Shunted Junction (RSJ) model (see Fig. 6). Extrapolating the $R_N^{-1}$ line at high voltages towards low voltages the intersection with 0 µV is not at zero current, but at higher values. This excess from 0 µA lowers the effective critical current dramatically. In the negative current branch of the graph in Fig. 6 the calculated $I_C$ value was 950 µA. But after subtracting the excess current of $I_{ex}$ = 645 µA the effective value of the critical current is as low as $I_C^{eff}$ = 305 µA.

Thus, the area independent junction parameter $I_C R_N$ is also decreased from 20.2 µV (formal value) to 6.5 µV (effective value) with a $R_N$ value of 21.3 mΩ.

This high amount of excess current was also noticed in measurements of the hybrid junction types. The planar SNS' junction formed a Josephson junction with a formal $I_C$ value of 350 µA (see Fig. 7). An excess current of about 200 µA lowered the $I_C R_N$ product from a formal value of 18.4 µV to the effective one of 7.9 µV [8]. We want to note again, that charge carriers are transported along the *c*-axis of the pnictide in this planar geometry, while they are transported within the *ab*-plane at both GB and edge-type geometry.

For edge-type Josephson junctions we found an excess current of about 24.0 µA (formal $I_C$ 58.8 µA; see Fig. 8). The resulting $I_C R_N$ product is reduced from 12.6 µV (formal) to 7.5 µV (effective) at this junction [9].

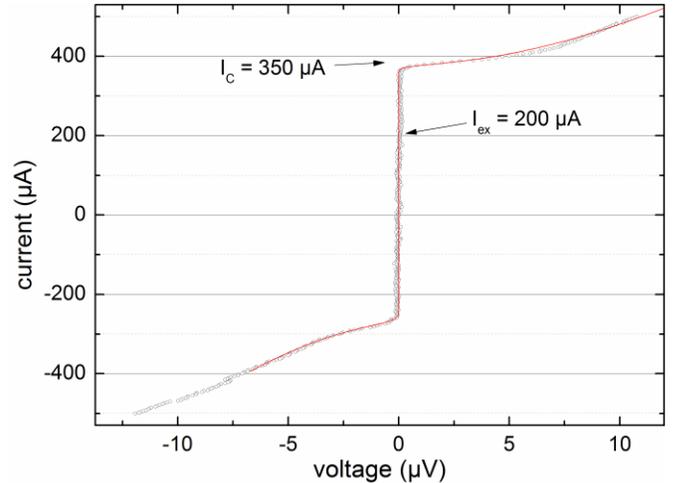

Fig. 7. Current-voltage characteristics of a planar SNS' junction at 4.2 K with Co-doped Ba-122 as base electrode and Pb as counter electrode, separated by an Au barrier of 5 nm thickness. The thin red line shows an asymmetric RSJ fit ($I_C^{pos}$=350 µA, $I_C^{neg}$=280 µA, $R_N^{neg}$=53.8 mΩ).

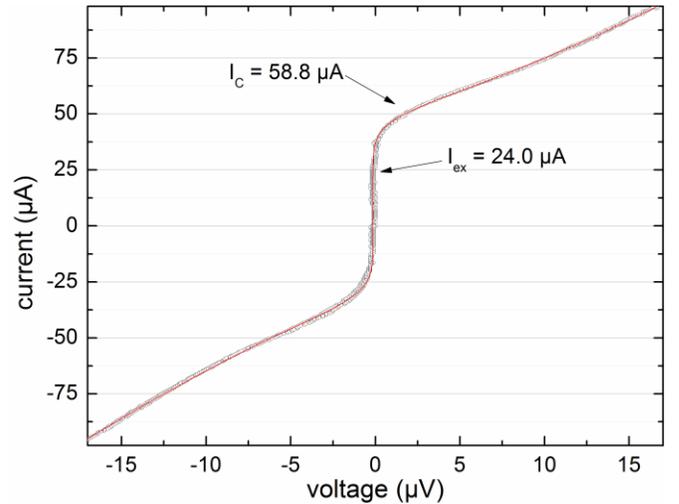

Fig. 8. Current-voltage characteristics of an edge-type junction at 4.2 K with Co-doped Ba-122 as base electrode and Pb as counter electrode measured in four-point geometry. The thin red line shows an asymmetric RSJ fit ($I_C^{pos}$=58.8 µA, $R_N^{pos}$=215 mΩ, $I_C^{neg}$=41.8 µA, $R_N^{neg}$=197 mΩ).

#### B. Influence of External Magnetic Fields

Exposure to magnetic fields leads to a periodically suppressed $I_C$ of the GB junction. Contrary to the theoretically calculated Fraunhofer pattern the amplitude reaches its maximum at a value different from zero magnetic field (see Fig. 9). Here the entire Fraunhofer-similar pattern is shifted towards negative coil currents. This shift occurs mainly due to an additional magnetic influence caused by the iron buffer layer (described in section II.A.) beneath the pnictide superconductor across the whole sample, especially in the junction region. An increase of the temperature above $T_C$ does not remove the trapped flux in the system due to the ferromagnetic influence of the buffer layer. In the planar junction geometry the Fe buffer layer is parallel to the barrier, so only stray fields of the ferromagnet influence the junction. In the edge-type junction the Fe buffer layer was removed during the IBE process together with the pnictide and thus has no influence.

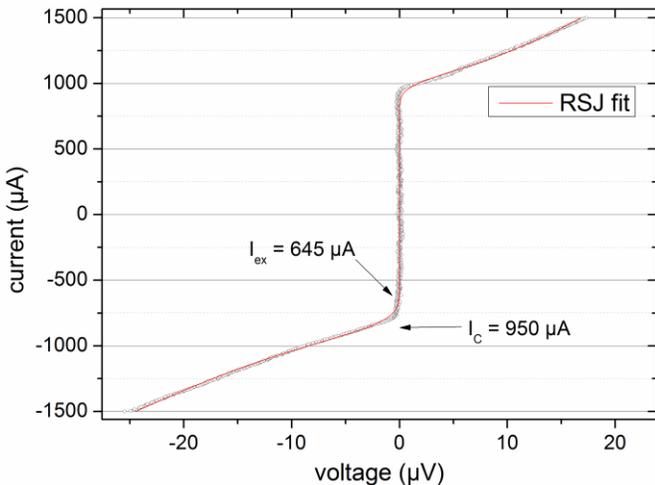

Fig. 6. Current-voltage characteristics of a grain boundary junction (bridge width 7 µm) at 4.2 K measured in four-point geometry. The thin red line shows an asymmetric RSJ fit ($I_C^{pos}$=1085 µA, $R_N^{pos}$=16.5 mΩ, $I_C^{neg}$=950 µA, $R_N^{neg}$=21.3 mΩ).

Additionally, the finite excess current (see section III.A) of the GB junction is present. It is influenced by the magnetic field in general, but parts of $I_{ex}$ cannot be suppressed entirely,

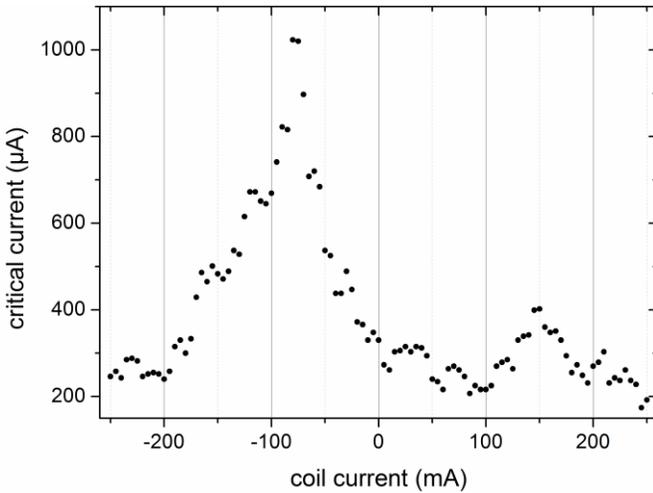

Fig. 9. Critical current of a GB junction vs. magnetic field (proportional to coil current) based on current-voltage characteristics at 4.2 K. This data was recorded before removal of the Au shunt layer.

leaving behind a current of about 200 µA.

Also, superposed local maxima with different periodicities can be seen. This effect is probably caused by a network of parallel Josephson junctions with slightly different properties within one bridge. Additionally, the pattern indicates that the contact is in the so called "long junction limit" [10], which can be avoided using narrower bridges.

*C. Temperature Dependence*

Measurements at temperatures between the boiling point of helium (4.2 K) and the critical temperature of the thin films (16.0 K) after the Au shunt removal (see section II.B.) were performed. In this temperature range, a nearly linear decay of $I_C$ can be seen when increasing the temperature. This behavior is similar to cuprate superconductor results [1] and the results of the planar SNS' and edge-type junctions [8], [9], where the temperature dependence is limited by the critical temperature of lead. The small $I_C$ value in Fig. 10 can be justified, because the critical current density was reduced from $1.4 \cdot 10^5$ A cm$^{-2}$ to $1.2 \cdot 10^4$ A cm$^{-2}$ at 4.2 K during the Au removal process. The reason for this strong decrease is not completely understood since destruction of a major part of the junction can be ruled out by consideration of the etching parameters. Nevertheless, the temperature dependence behaved similar before the additional etching process.

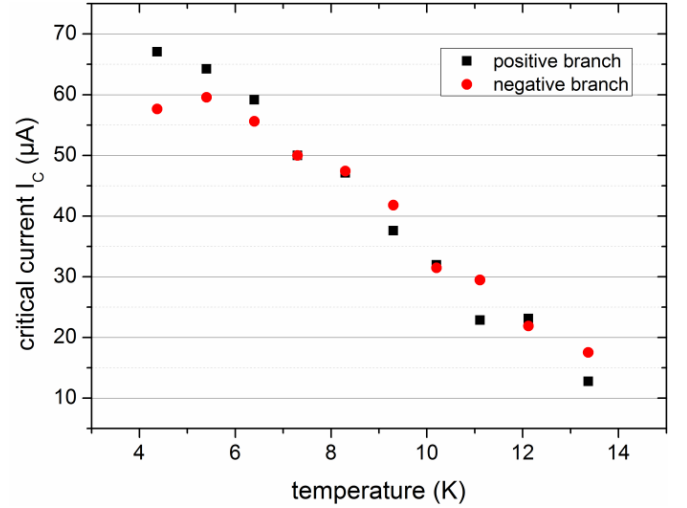

Fig. 10. Critical current of a GB junction vs. temperature. A slight asymmetry is visible at low temperatures between positive current branch and negative current branch of the I-V characteristics taken as a basis for this graph.

## IV. CONCLUSION

The successful realization of three entirely different thin film junction geometries allows us to understand and investigate effects, which originate from the examined material (i.e. Co-doped Ba-122). In all three geometries we found a high amount of excess current within the formal critical current. This leads to the assumption that $I_{ex}$ is not caused by the junctions or the different types of barriers, but by the pnictide thin film itself. Furthermore, we showed that an iron buffer is crystallographically preferable, but influences the magnetic field response of the junction greatly. Controlling the magnetic properties of the Fe buffer opens a path for further device oriented developments for iron-based superconductors.

ACKNOWLEDGMENT

We like to thank I. Mönch and V. Grosse for their contributions.